\documentclass[conference]{IEEEtran}

\usepackage{cite}
\usepackage{graphicx}
\usepackage{epstopdf}
\usepackage{gensymb,bm}
\usepackage[T1]{fontenc}
\usepackage[utf8]{inputenc}
\usepackage{textcomp}
\usepackage[cmex10]{amsmath}
\usepackage{amssymb}
\usepackage{booktabs}
\usepackage{url}
\usepackage{subcaption}
\usepackage{multirow}
\usepackage{adjustbox}
\usepackage{siunitx}

\begin{document}

\title{Ultra-wideband Double-Directionally Resolved Channel Measurements of Line-of-Sight Microcellular Scenarios in the Upper Mid-band}

\author{\IEEEauthorblockN{Naveed A. Abbasi\IEEEauthorrefmark{1}\IEEEmembership{Member,~IEEE}, Kelvin Arana\IEEEauthorrefmark{1}, Jorge Gomez-Ponce\IEEEauthorrefmark{1}\IEEEauthorrefmark{2}\IEEEmembership{Member,~IEEE,}, Tathagat Pal\IEEEauthorrefmark{1}, Vikram Vasudevan\IEEEauthorrefmark{1}, Atulya Bist\IEEEauthorrefmark{1},\\Omer Gokalp Serbetci \IEEEauthorrefmark{1}, Young Han Nam\IEEEauthorrefmark{3}, Charlie Zhang\IEEEauthorrefmark{3}\IEEEmembership{Fellow,~IEEE}, Andreas F. Molisch\IEEEauthorrefmark{1}}\IEEEmembership{Fellow,~IEEE}
 \IEEEauthorblockA{\IEEEauthorrefmark{1}University of Southern California, Los Angeles, CA, USA}
\IEEEauthorblockA{\IEEEauthorrefmark{2}ESPOL Polytechnic University, Guayaquil, Ecuador\\}
\IEEEauthorblockA{\IEEEauthorrefmark{3}Samsung Research America, Richardson, TX, USA} 
	Email: \{nabbasi, aranaoro, gomezpon, tpal, vikramva, bist, serbetci, molisch\}@usc.edu,\\ \{younghan.n, jianzhong.z\}@samsung.com}

\maketitle

\begin{abstract}
The growing demand for higher data rates and expanded bandwidth is driving the exploration of new frequency ranges, including the upper mid-band spectrum (6-24 GHz), which is a promising candidate for future Frequency Range 3 (FR3) applications. This paper presents ultra-wideband double-directional channel measurements in line-of-sight microcellular scenarios within the upper mid-band spectrum (6-18 GHz). Conducted in an urban street canyon environment, these measurements explore key channel characteristics such as power delay profiles, angular power spectra, path loss, delay spread, and angular spread to provide insights essential for robust communication system design. Our results reveal that path loss values for both omni-directional and best beam configurations are lower than free-space predictions due to multipath contributions from the environment. Analysis also indicates a high degree of stability in delay spread and angular spread across the entire band, with small variation between sub-bands.
\end{abstract}


\section{Introduction}
The demand for high data rates in urban environments is increasingly driving the exploration of the upper mid-band spectrum. This spectrum, considered as a potential Frequency Range 3 (FR3), offers promising opportunities by potentially enabling new levels of connectivity and coverage \cite{kang2024cellular}. However, robust channel characterization based on precision measurements within this frequency range is essential to support effective communication system design.

In particular, as multiple-input multiple-output (MIMO) technologies are expected to play a key role in achieving the desired performance, the measurements should ideally be double-directional (i.e., directionally resolved at both link ends) to accurately capture both the transmit and receive characteristics \cite{bjornsonGiganticMIMO}. Given the ongoing uncertainty surrounding the specific bands that will ultimately be allocated for such high-data-rate applications, ultra-wideband (UWB) measurements are also necessary to ensure the flexibility and comprehensiveness of the data collected. Furthermore, it is important to understand the frequency dependence of channel model parameters, as specified in frameworks like 3GPP, to ensure compatibility and adaptability across potential operational bands \cite{3gppFR3}.



There have been a number of propagation measurements in the upper mid-band,\footnote{There are a number of measurements in the FCC-defined UWB band 3.1-10.6 GHz, which partly overlaps with the upper mid-band. We do not consider such measurements here and refer to \cite{molisch2009ultra} for a literature survey.} which can be categorized as follows:\footnote{Due to space constraints, only example papers can be given.} (i) measurement of specific propagation effects, such as the transmission and reflection from certain building materials \cite{shakya2024wideband}, diffraction \cite{tervo2014diffraction}, and presence of people \cite{janssen1996wideband}; (ii) measurements where both link ends use only a single (typically omni-directional antenna; those measurements can be narrowband, e.g., \cite{oh2019empirical,yoza2019path}, wideband \cite{salous2014radio}, or UWB \footnote{We define here UWB as having more than 20 \% relative bandwidth, in line with one of the definitions of the FCC. While the FCC also allows to classify any transmission with more than 500 MHz bandwidth as UWB, this definition mainly stems from regulatory considerations of spectrum spreading, but does not relate easily to variations of the channel statistics over the bandwidth \cite{molisch2009ultra}. We furthermore categorize measurements that are measuring multiple, widely separated sub-bands, where each of them is wideband, as wideband.} \cite{kristem2018outdoor}; (iii) directional at one link end, which may be wideband \cite{fan2016measured,naderpour2016spatio} or UWB \cite{ling2016comparison,wang2022propagation,miao2023sub}; and (iv) double-directional measurements \cite{saito2017dense,kurita2016indoor, roivainen2016validation}, which are all wideband (up to 1 GHz in \cite{shakya2024urban}), but not UWB.  

Building on the previous discussion, \textit{this paper presents the initial results from the first double-directional UWB measurement campaign in the upper mid-band}. Specifically, we analyze the 6-18 GHz range in an urban line-of-sight (LoS) microcellular scenario, based on over 14,000 directional power delay profiles (PDPs). In this setup, the transmitter (Tx) is positioned over 20 meters above ground, with the receivers (Rx) located at distances ranging from 60 to 185 meters. We examine both the full band and 1 GHz sub-bands to reveal key channel characteristics, including path loss, delay spread, and angular spread, both as a function of distance and as a function of frequency. These findings lay the groundwork for reliable upper mid-band channel models in LoS conditions. It is important to note that our transmitter height of over 20 meters is close to the 25-meter standard for Urban Macro (UMa) scenarios in 3GPP with our position atop a building. However, since we cover a sector on the base station, we use the Urban Micro (UMi) notation. Nevertheless, the results obtained from this measurement campaign may also provide valuable insights for UMa analysis.

\section{Measurement Setup and Site}
\subsection{Testbed description}

For this measurement campaign, we use a custom RF-over-fiber (RFoF) frequency-domain channel sounder, similar to \cite{abbasi2022thz,abbasi2023thz}, featuring a vector network analyzer (VNA) covering a 12 GHz bandwidth between 6 and 18 GHz with HGHA618 high-gain horn antennas (6-18 GHz) as the front-end. 
In this microcellular scenario, the Tx is mounted on a building over 20 meters high, while the Rx, representing a user equipment (UE), is positioned on the ground at 1.7 meters above ground. Each VNA sweep captures 12,001 frequency points across the 8 GHz bandwidth, enabling measurements of excess delays up to 1 $\mu$s, corresponding to a maximum multipath run length of 300 meters - sufficient for the scenario and frequency band under study.

Precision mechanical positioners rotate the antennas in both azimuth and elevation, enabling double-directional measurements, i.e., determination of the complex transfer function as a function of direction of arrival (DoA), and direction of departure (DoD). Each double-directional measurement takes several hours, mainly due to the long aggregate time to rotate the antennas to their various angles, and settling time to ensure that the antenna does not vibrate during the actual measurement. Consequently, measurements were conducted at night to maintain a static environment. Potential movement from pedestrians was avoided by access restrictions to the measurement area; however, varying wind speeds may affect vegetation during the measurement.

Table \ref{table:parameters} outlines the measurement parameters: the Tx is fixed at one elevation angle, covering a sector from $-60^\circ$ to $60^\circ$ in $10^\circ$ increments, while the Rx spans five co-elevations from $-20^\circ$ to $20^\circ$ with a $10^\circ$ stepwidth and $0^\circ$ to $360^\circ$ azimuth coverage in $10^\circ$ steps. Consequently, each Tx/Rx location pair includes $2340$ transfer functions. 
The setup aligns the Tx co-elevation and Rx $0^\circ$ co-elevation ($\tilde{\theta}_{Rx}$) so that the LOS MPC has a co-elevation of $0^\circ$. To isolate "system and antenna" effects from the measurements, we conduct a time-gated over-the-air (OTA) calibration daily throughout the campaign at a point 44 meters away from the Tx (as shown in Fig. \ref{fig:site}).  

\begin{table}[t!]
	\centering
 \vspace{1mm}
	\caption{Setup parameters.}
	\label{table:parameters}
	\begin{tabular}{|l|c|l|}
		\hline
		\textbf{Parameter}              & \textbf{Symbol}   & \textbf{Value} \\ \hline\hline
		\textit{Frequency range}        & $f$                 & 6-18 GHz           \\
		\textit{Tx height}              & $h_{Tx}$           & $20.3m$               \\
		\textit{Rx height}              & $h_{Rx}$           & $1.7m$               \\
		\textit{Elevation, Azimuth resolution} & $\Delta\phi_{Tx}$, $\Delta\phi_{Rx}$, $\Delta\theta_{Rx}$ & 10$^{\circ}$ \\
		\textit{Tx Azimuth rotation range}   & $\phi_{Tx}$        & [-60$^{\circ}$,60$^{\circ}$]           \\
		\textit{Rx Azimuth rotation range}   & $\phi_{Rx}$        & [0$^{\circ}$,360$^{\circ}$]           \\
		\textit{Rx Elevation rotation range}   & $\theta_{Rx}$        & [-20$^{\circ}$,20$^{\circ}$]   \\        
		\hline
	\end{tabular}
\end{table}

\subsection{Measurement site}
\begin{figure}[t!]
	\centering
 \vspace{1mm}
	\includegraphics[width=5 cm]{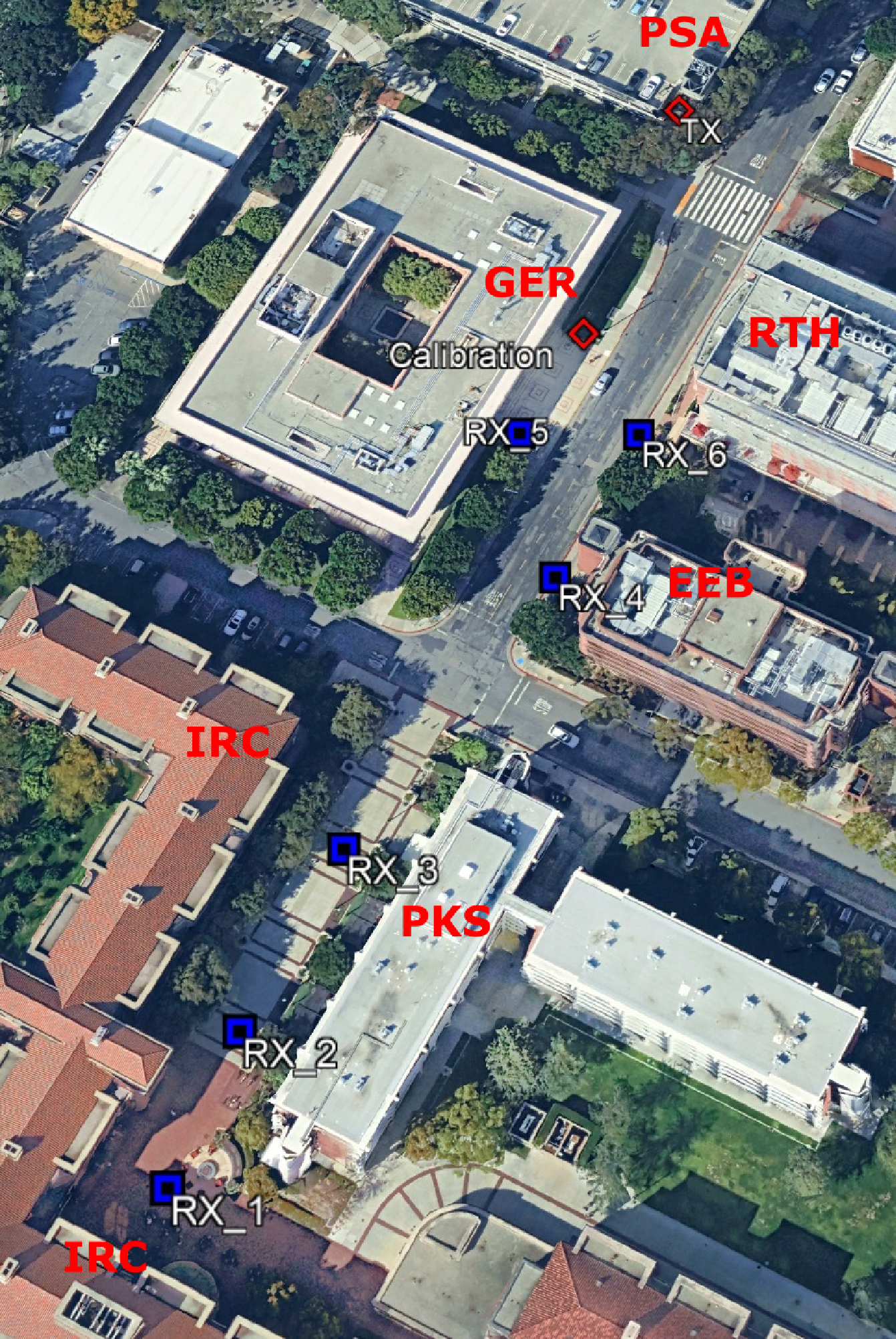}
	\caption{Measurement site.}
	\label{fig:site}%

\end{figure}
The measurements are conducted along McClintock Avenue on the University of Southern California's University Park Campus (UPC) in Los Angeles, California, USA. The Tx is positioned near the top of the Downey Way Parking Structure (PSA) on an external staircase, providing an elevated line-of-sight vantage point for signal transmission across the measurement area. The Rx points are distributed along McClintock Avenue, extending from Rx1 to Rx6, which are spread at varying distances from the Tx to capture a range of channel characteristics. The farthest Rx point, Rx1, is located near the IRC Building. The overall environment is an urban street canyon with a blend of building facades, parked cars, and foliage, providing a realistic setting for studying channel characteristics in an urban microcellular scenario.

A detailed map showing the locations of Tx and Rx points is presented in Fig. \ref{fig:site} and the details of the points and their distances from Tx are given in Table \ref{table:rx_points}.

\begin{table}[h!]
\centering
\vspace{-3mm}
\caption{Rx points with corresponding distances from Tx.}
\label{table:rx_points}
\begin{adjustbox}{width=8.5cm}
\begin{tabular}{|c||c|c|c|c|c|c|}
\hline
\textbf{Rx Point} & \textbf{Rx1} & \textbf{Rx2} & \textbf{Rx3} & \textbf{Rx4} & \textbf{Rx5} & \textbf{Rx6} \\ \hline \hline
\textbf{Distance from Tx (m)} & 184.8 & 161.7 & 132.6 & 83.4 & 63.6 & 59.4 \\ \hline
\end{tabular}
\end{adjustbox}
\vspace{-3mm}
\end{table}

\section{Parameters and Processing}
The VNA-based measurement setup produces frequency scans at various Tx-Rx locations, forming a four-dimensional tensor $H_{meas}(f,\phi_{Tx},\phi_{Rx},\tilde{\theta}_{Rx};d)$, where $f$ is the frequency, $\phi_{Tx}$ and $\phi_{Rx}$ represent Tx and Rx azimuth orientations,  $\tilde{\theta}_{Rx}$ denotes the  Rx elevation, and $d$ is the Tx-Rx distance. The calibrated channel transfer function is obtained by dividing $H_{meas}$ by the OTA calibration, $H_{OTA}(f)$:
\begin{align}
    H(f,\phi_{Tx},\phi_{Rx},\tilde{\theta}_{Rx};d) &= 
    \frac{H_{meas}(f,\phi_{Tx},\phi_{Rx},\tilde{\theta}_{Rx};d)}
    {H_{OTA}(f)}.
\end{align}

The directional PDP is then computed as:
\begin{multline}
    P_{calc}(\tau,\phi_{Tx},\phi_{Rx},\tilde{\theta}_{Rx};d) = \\ 
    \left| \mathcal{F}_{f}^{-1} \left\{ H(f,\phi_{Tx},\phi_{Rx},\tilde{\theta}_{Rx};d) \right\} 
    \right|^2.    
\end{multline}

Noise reduction is achieved using thresholding and delay gating similar to \cite{abbasi2022thz,abbasi2023thz}, defined as:
\begin{align}
    P(\tau) &= \left[ P_{calc}(\tau) : (\tau \leq \tau_{gate}) \land 
    (P_{calc}(\tau) \geq P_{\lambda}) \right],
\end{align}
where $\tau_{gate}$ and $P_{\lambda}$ are set based on noise characteristics. We refer to \cite{gomez2023impact} concerning the impact of the selection of noise and delay thresholds.

The strongest directional PDP (Max-Dir) is selected as the beam-pair with the highest power:
\begin{align}
    P_{\rm Max-Dir}(\tau;d)
     &= \arg\max_{i,j,k} \sum_\tau 
    P(\tau,\phi_i,\phi_j,\tilde{\theta}_k;d).
\end{align}

An "omni-directional" PDP is constructed by combining co-elevations and selecting the strongest azimuth direction per delay bin similar to \cite{Hur_omni}:
\begin{align}
    P_{\rm omni}(\tau;d) &= \max_{\phi_{Tx},\phi_{Rx}} \sum_i 
    P(\tau,\phi_{Tx},\phi_{Rx},\tilde{\theta}_{Rx}^i;d),
\end{align}
where $i \in \{1, 2, 3, 4, 5\}$ represents Rx co-elevations, spaced $10^\circ$ apart. 
\begin{figure}[t!]
	\centering
	\includegraphics[width=0.75\linewidth]{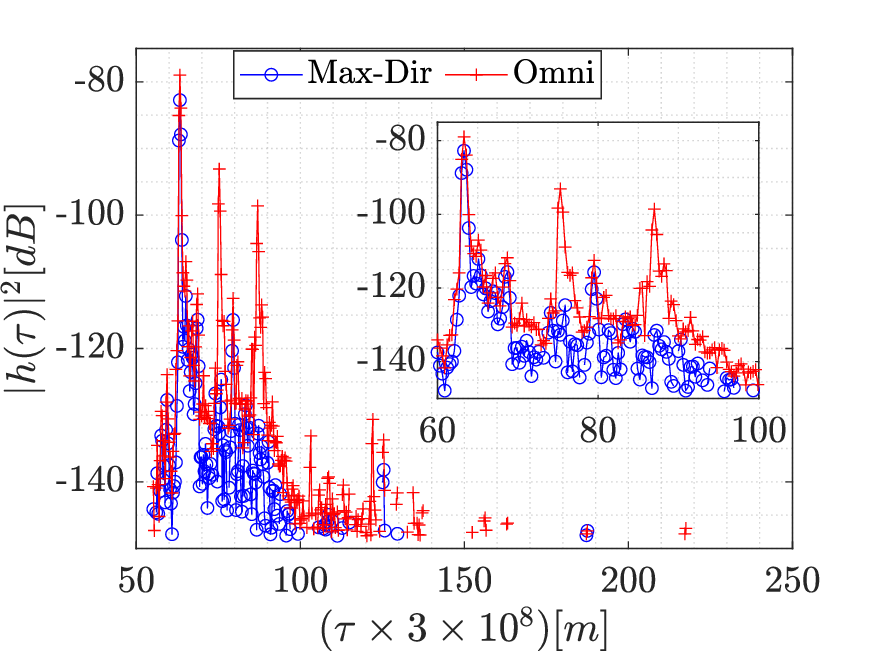}
	\caption{PDP for Rx5 at $63.6m$.}
	\label{fig:PDP}%
 \vspace{-5mm}
\end{figure}
We refer to \cite{abbasi2023thz, abbasi2022thz} for parameter definitions, which include: path gain ($PG$), derived from the sum of power across all delay bins in a PDP; path loss ($PL$), calculated as the inverse of $PG$ and modeled with a power-law relationship using parameters $\alpha$ and $\beta$, with shadowing represented as $N(0, \sigma)$ \cite{molisch2023wireless}; the root mean square delay spread (RMSDS, $\sigma_\tau$), representing the second central moment of the PDP \cite{molisch2023wireless}; and the angular spread (AS, $\sigma^\circ$), determined using Fleury’s definition based on the double-directional angular power spectrum (DDAPS), which provides a measure of the angular dispersion at either the Tx or Rx \cite{fleury2000first, molisch2023wireless}. To see the relative statistics of RMSDS and AS on a sub-band level, we performed a fitting of the cumulative density function (CDF) of the measured RMSDS and AS with a normal (Gaussian) distribution.

\section{Measurement Results}


\subsection{Power delay profile}

Fig. \ref{fig:PDP} shows a sample PDP, namely at location Rx5. Note that the delay on the x-axis is given in meters, which corresponds to the runlength (calculated as delay \(\times\) \(c\), where \(c\) is the speed of light, \(3 \times 10^8\) $m/s$) since this transformation facilitates a more intuitive analysis.
\begin{figure}[t!]
	\centering
	\includegraphics[width=0.75\linewidth]{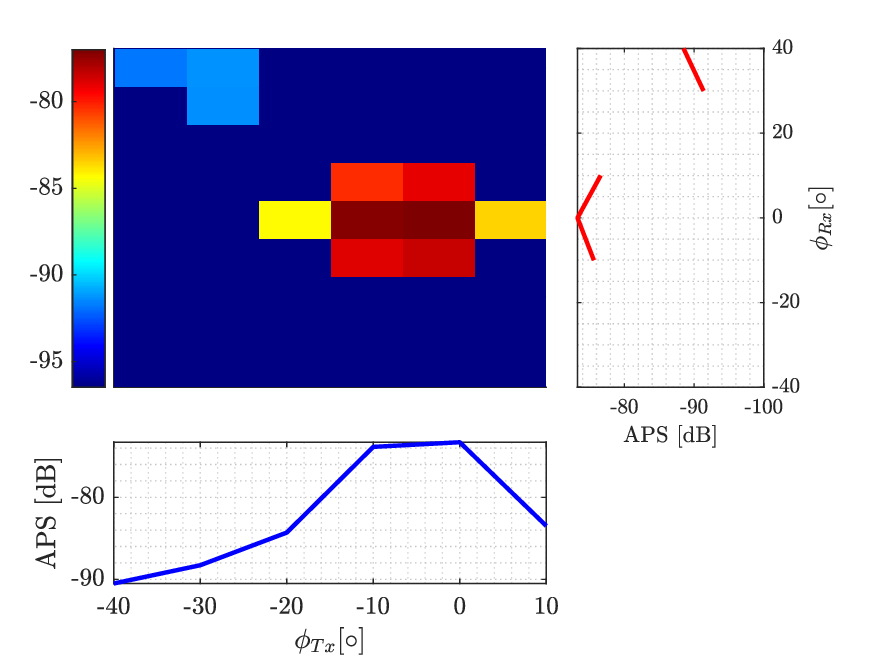}
	\caption{APS for Rx5 at $63.6m$.}
	\label{fig:APS}%
 \vspace{-6mm}
\end{figure}
Here, we see that the LoS component is present at a delay corresponding to $63.3m$, which aligns with the physical Tx-Rx distance. The omni-directional PDP exhibits more multipath components (MPCs) compared to the Max-Dir PDP, as it receives additional MPCs from multiple directions, including reflections from buildings across the road which represent the strongest reflections. Apart from these, a number of smaller MPCs are seen, with a few having delays beyond $200m$. Despite their lower power relative to the strongest components, these additional reflections increase the delay spread in the omni-directional PDP. In contrast, these components are notably diminished in the Max-Dir profile due to the spatial filtering effect provided by the directional antennas. Consequently, the Max-Dir profile presents a narrower range of DoAs and DoDs, which has implications for delay spread and angular spread.
\subsection{Angular power spectrum}

\begin{figure*}[t!]
    \begin{subfigure}[b]{0.32\textwidth}
    \centering
        \includegraphics[width=1\columnwidth]{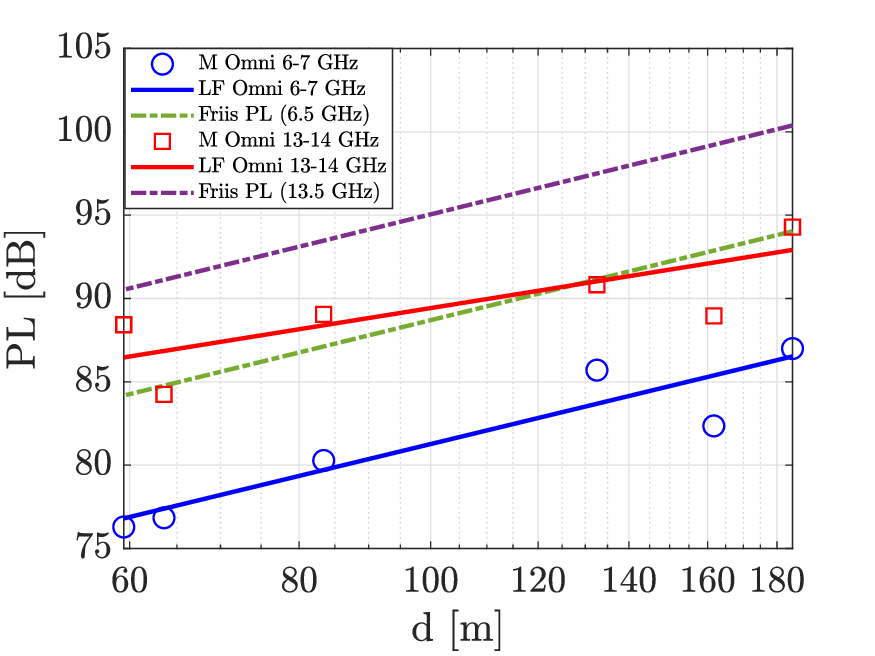}
        \caption{Omni Path loss modeling.}
        \vspace*{0mm}
        \label{PLOSS-Omni}
        \end{subfigure}
    \begin{subfigure}[b]{0.32\textwidth}
    \centering
        \includegraphics[width=1\columnwidth]{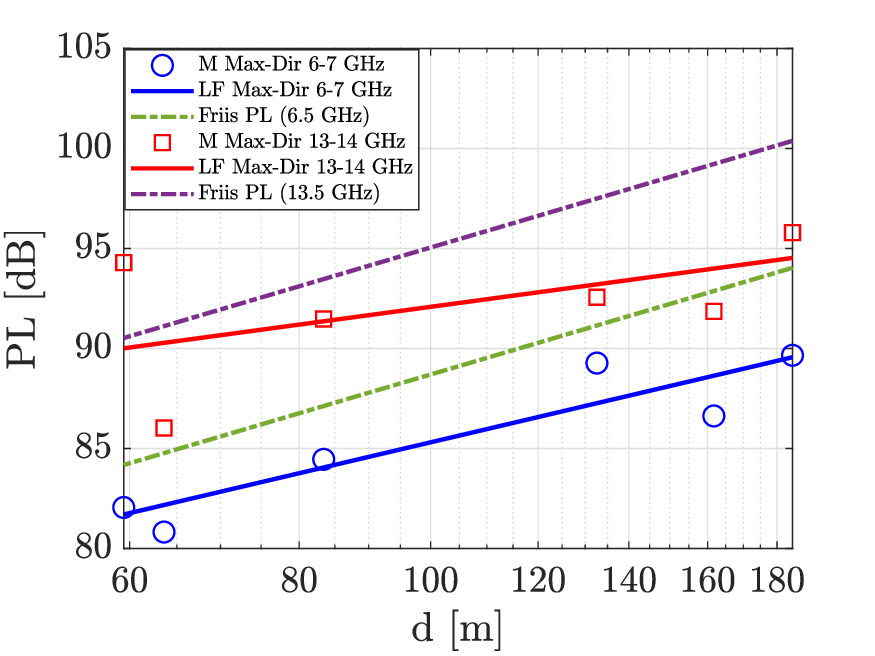}
        \caption{Max-Dir Path loss modeling.}
        \vspace*{0mm}
        \label{PLOSS-Max-Dir}
        \end{subfigure}
    \centering
    \begin{subfigure}[b]{0.32\textwidth}
        \centering
        \includegraphics[width=1\columnwidth]{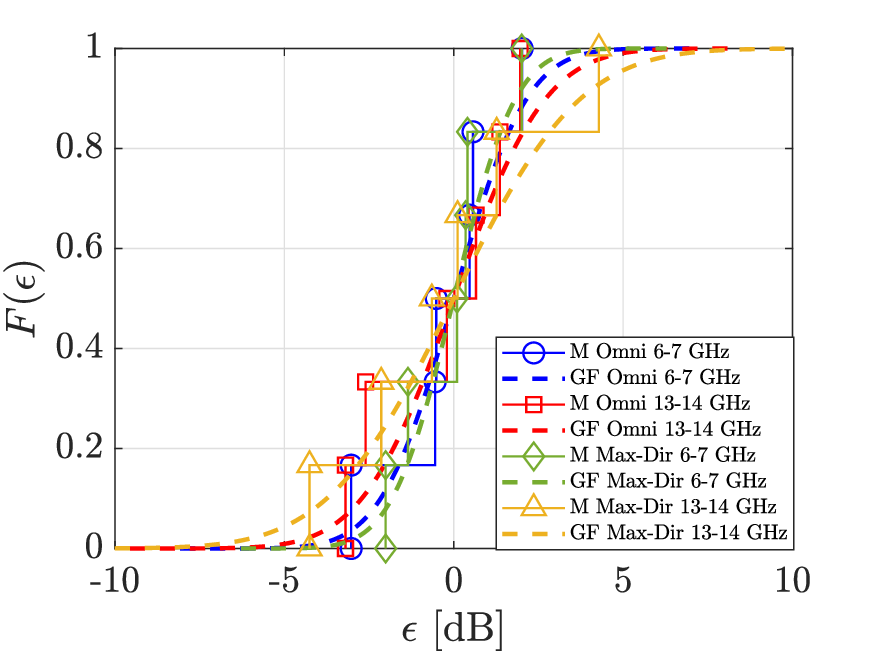}
        \caption{Shadowing.}
        \vspace*{0mm}
        \label{SHA-LOS}
    \end{subfigure}
\caption{Path loss and shadowing models for LoS points. Measured (M), Gaussian Fit (GF), and Linear Fit (LF) curves.}%
\label{fig:los_PL_SHA}%
\vspace{-0 mm}
\vspace{-5mm}
\end{figure*}
For the analysis of the angular characteristics, the resolution of the capture is directly related to the angular step size of the rotors (see Table \ref{table:parameters}). Fig. \ref{fig:APS}  shows the angular power spectrum (APS) for the same measurement location, demonstrating that the power is concentrated around the LoS. Besides the LoS component, at the Rx side, a smaller cluster comes from an azimuth angle of $40^\circ$, i.e., facing the opposite of the street towards the Ronald Tutor Hall (RTH) building. On the Tx side, the energy is distributed from $-40^\circ$ to $10^\circ$ with the highest concentration from $-20^\circ$ to $10^\circ$. The negative angles are related to reflections hitting the RTH building on the opposite side of the street, while the positive angles are related to reflections hitting the GER building.

Mapping these directions to the peaks in the PDP shown in Fig. \ref{fig:PDP}, we observe the LoS matches with the strongest peak at $63m$, the second and third strongest peaks shown in the omnidirectional PDP at $75.3m$ and $87m$, respectively, correspond to reflections hitting RTH. The components between $63m$ and $75m$ correspond to the reflections at $10^\circ$ hitting GER building, and attenuated it is partially hitting the side of the building and reaching Rx5. This excellent agreement between the extracted values and the physics of the environment also indicates that the measurement results are reliable and accurate. 

\subsection{Path loss and shadowing}
The detailed band-by-band parameters for Omni path loss modeling, Max-Dir path loss modeling, and shadowing modeling are presented in Tables \ref{tab:lf_PG_Omni}, \ref{tab:lf_PG_Max-Dir}, and \ref{tab:shadowing}, respectively. We further highlight specific sub-bands in Fig. \ref{fig:los_PL_SHA} to discuss some notable observations.

The Omni path loss modeling in Fig. \ref{PLOSS-Omni} shows that path loss increases with both frequency and distance, as expected. We also note that each of the parameters $\alpha$ and $\beta$ might not increase over frequency, but that the path loss (over the range of observed distances) {\em does} show such an increase. Additionally, our results demonstrate lower path loss than is obtained in free space (Friis' equation), likely due to the MPCs captured within the omni-directional response. For Max-Dir path loss modeling, shown in Fig. \ref{PLOSS-Max-Dir}, the pathloss is generally higher (due to not collecting MPCs outside the main beam), but
the general observations remain consistent. A notable exception here is Rx6, which shows a higher path loss than predicted by the Friis model. This deviation is due to partial obstruction pf the LoS from branches of a nearby tree. This finding underscores the importance and necessity of performing detailed vegetation analysis at these frequencies, an area we will explore further in our future studies.

\begin{table}[t!]
\centering
\caption{Linear fitting for $PG_{Omni}$ with 95$\%$ confidence intervals.}
\label{tab:lf_PG_Omni}
\begin{adjustbox}{width=8.5cm}
\begin{tabular}{|c||c|c|c|c|c|c|}
\hline
\textbf{Frequency} & $\boldsymbol{\alpha_{min,95\%}}$ & $\boldsymbol{\alpha}$ & $\boldsymbol{\alpha_{max,95\%}}$ & $\boldsymbol{\beta_{min,95\%}}$ & $\boldsymbol{\beta}$ & $\boldsymbol{\beta_{max,95\%}}$ \\ \hline \hline
All Bands & 40.60 & 57.97 & 75.34 & 0.65 & 1.50 & 2.35 \\ 
6-7 GHz & 21.59 & 41.82 & 62.05 & 0.98 & 1.97 & 2.96 \\ 
7-8 GHz & 36.43 & 55.24 & 74.05 & 0.49 & 1.41 & 2.33 \\ 
8-9 GHz & 29.95 & 47.12 & 64.29 & 1.07 & 1.91 & 2.75 \\ 
9-10 GHz & 34.16 & 56.55 & 78.95 & 0.31 & 1.40 & 2.50 \\ 
10-11 GHz & 33.93 & 55.43 & 76.92 & 0.57 & 1.63 & 2.68 \\ 
11-12 GHz & 30.72 & 50.54 & 70.36 & 0.96 & 1.93 & 2.90 \\ 
12-13 GHz & 38.14 & 59.35 & 80.56 & 0.47 & 1.51 & 2.55 \\ 
13-14 GHz & 36.78 & 63.24 & 89.69 & 0.02 & 1.31 & 2.60 \\ 
14-15 GHz & 52.86 & 70.70 & 88.54 & 0.03 & 0.90 & 1.77 \\ 
15-16 GHz & 53.69 & 73.42 & 93.16 & -0.21 & 0.76 & 1.72 \\ 
16-17 GHz & 52.76 & 72.90 & 93.04 & -0.24 & 0.75 & 1.73 \\ 
17-18 GHz & 39.31 & 64.36 & 89.41 & -0.04 & 1.19 & 2.41 \\ 
\hline
\end{tabular}
\end{adjustbox}
\vspace{-3mm}
\end{table}

\begin{table}[t!]
\centering
\caption{Linear fitting for $PG_{Max-Dir}$ with 95$\%$ confidence intervals.}
\label{tab:lf_PG_Max-Dir}
\begin{adjustbox}{width=8.5cm}
\begin{tabular}{|c||c|c|c|c|c|c|}
\hline
\textbf{Frequency} & $\boldsymbol{\alpha_{min,95\%}}$ & $\boldsymbol{\alpha}$ & $\boldsymbol{\alpha_{max,95\%}}$ & $\boldsymbol{\beta_{min,95\%}}$ & $\boldsymbol{\beta}$ & $\boldsymbol{\beta_{max,95\%}}$ \\ \hline \hline
All Bands & 48.08 & 68.09 & 88.09 & 0.17 & 1.15 & 2.13 \\ 
6-7 GHz & 36.52 & 53.41 & 70.29 & 0.77 & 1.60 & 2.42 \\ 
7-8 GHz & 51.22 & 64.96 & 78.70 & 0.44 & 1.11 & 1.79 \\ 
8-9 GHz & 43.73 & 56.81 & 69.90 & 0.96 & 1.60 & 2.24 \\ 
9-10 GHz & 46.12 & 66.07 & 86.01 & 0.12 & 1.09 & 2.07 \\ 
10-11 GHz & 42.51 & 63.47 & 84.42 & 0.36 & 1.39 & 2.41 \\ 
11-12 GHz & 38.74 & 58.03 & 77.32 & 0.77 & 1.71 & 2.65 \\ 
12-13 GHz & 38.12 & 68.31 & 98.50 & -0.28 & 1.20 & 2.68 \\ 
13-14 GHz & 38.69 & 73.75 & 108.81 & -0.80 & 0.92 & 2.63 \\ 
14-15 GHz & 53.97 & 81.20 & 108.43 & -0.83 & 0.50 & 1.84 \\ 
15-16 GHz & 52.80 & 81.83 & 110.86 & -0.97 & 0.45 & 1.87 \\ 
16-17 GHz & 55.15 & 81.94 & 108.73 & -0.88 & 0.43 & 1.74 \\ 
17-18 GHz & 41.46 & 73.83 & 106.20 & -0.76 & 0.82 & 2.41 \\ 
\hline
\end{tabular}
\end{adjustbox}
\vspace{-3mm}
\end{table}

\begin{table}[t!]
\centering
\caption{Shadowing distribution parameters with 95$\%$ confidence intervals.}
\label{tab:shadowing}
\begin{adjustbox}{width=8.5cm}
\begin{tabular}{|c||c|c|c|c|c|c|}
\hline
\multirow{2}{*}{\textbf{Frequency}} & \multicolumn{3}{c|}{$\boldsymbol{PG_{Omni}}$} & \multicolumn{3}{c|}{$\boldsymbol{PG_{Max-Dir}}$} \\ \cline{2-7} 
& $\boldsymbol{\sigma}$ & $\boldsymbol{\sigma_{\text{min, 95\%}}}$ & $\boldsymbol{\sigma_{\text{max, 95\%}}}$ & $\boldsymbol{\sigma}$ & $\boldsymbol{\sigma_{\text{min, 95\%}}}$ & $\boldsymbol{\sigma_{\text{max, 95\%}}}$ \\ \hline \hline
All Bands & 1.37 & 0.42 & 1.70 & 1.66 & 0.70 & 2.15 \\ 
6-7 GHz & 1.68 & 0.52 & 2.29 & 1.42 & 0.68 & 1.83 \\ 
7-8 GHz & 1.52 & 0.55 & 1.96 & 1.09 & 0.49 & 1.28 \\ 
8-9 GHz & 1.28 & 0.57 & 1.73 & 1.01 & 0.31 & 1.11 \\ 
9-10 GHz & 1.77 & 0.36 & 2.39 & 1.61 & 0.36 & 1.95 \\ 
10-11 GHz & 1.67 & 0.50 & 2.25 & 1.68 & 0.79 & 2.07 \\ 
11-12 GHz & 1.50 & 0.56 & 1.80 & 1.55 & 0.68 & 1.90 \\ 
12-13 GHz & 1.70 & 0.76 & 2.11 & 2.51 & 0.91 & 3.37 \\ 
13-14 GHz & 2.12 & 0.64 & 2.52 & 2.92 & 1.19 & 3.92 \\ 
14-15 GHz & 1.50 & 0.51 & 1.91 & 2.32 & 0.26 & 3.26 \\ 
15-16 GHz & 1.65 & 0.69 & 1.98 & 2.42 & 0.86 & 3.25 \\ 
16-17 GHz & 1.69 & 0.74 & 2.00 & 2.25 & 0.39 & 2.99 \\ 
17-18 GHz & 2.04 & 1.04 & 2.57 & 2.63 & 1.17 & 3.26 \\ \hline
\end{tabular}
\end{adjustbox}
\vspace{-5mm}
\end{table}
The shadowing distributions illustrated in Fig. \ref{SHA-LOS} reveal similar curves across Max-Dir and Omni configurations, and across frequencies, with a small variation in the standard deviation (\(\sigma\)) of the zero-mean lognormal fit at different frequencies. This difference is likely due to the frequency selectivity in the channel and the decreasing dynamic range as the frequency increases. Finally, we note that the overall results in Tables \ref{tab:lf_PG_Omni}, \ref{tab:lf_PG_Max-Dir}, and \ref{tab:shadowing} show that the 95\% confidence interval for minimum and maximum values is quite broad in certain cases, though with the same sign, highlighting the need for a larger sample size to better characterize the conditions and increase the precision of the model; our future work will thus include additional measurement campaigns.  
\subsection{RMS delay spread}
\begin{figure}[t!]
	\centering
	\includegraphics[width=0.75\linewidth]{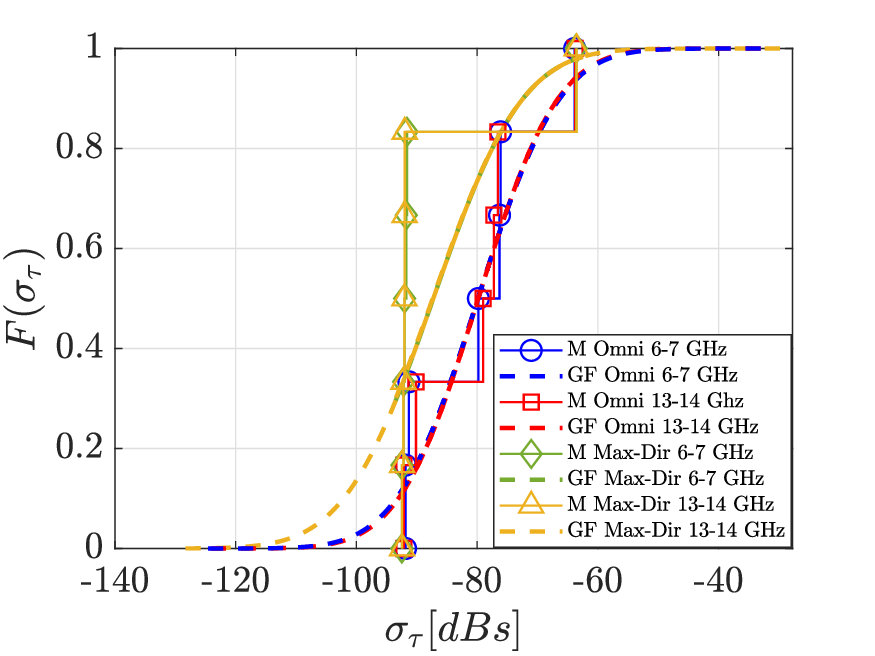}
	\caption{Modeling of RMS delay spread. Measured (M), and Gaussian Fit (GF) curves.}
	\label{fig:RMSDS}%
 \vspace{-7mm}
\end{figure}

Next, we discuss RMSDS, which is analyzed on a logarithmic scale, i.e., \SI{}{\decibel\second} ($10 \log(\text{Delay Spread}/1~\text{second})$), as is common in the channel modeling literature, such as in 3GPP. Results across all sub-bands are presented in Tables \ref{tab:nf_RMDDS_Omni} and \ref{tab:nf_RMSDS_Max-Dir} for normal fitting of the CDF. We also performed linear fitting where we observed significant variation in delay spread with distance, though the confidence intervals in both the Omni or Max-Dir cases encompassed positive and negative values for the slope so they are not discussed further (and not shown due to space limitations).

The delay spread distributions provide additional insights. A sample comparison of RMSDS for Max-Dir and Omni at the 6-7 GHz and 13-14 GHz bands (Fig. \ref{fig:RMSDS}) reveals that RMSDS for Max-Dir is lower than in the Omni cases, which we attribute to the spatial filtering effect of the antenna in the best-beam case. For Max-Dir, both the 6-7 GHz (green) and 13-14 GHz (yellow) bands show a high degree of overlap, with RMSDS values centered around -87 \SI{}{\decibel\second}. The distributions across various frequencies are quite similar for the Omni case as well, with an Omni-directional mean delay spread close to -80 \SI{}{\decibel\second}. This uniformity over frequency, further confirmed in Tables \ref{tab:nf_RMDDS_Omni} and \ref{tab:nf_RMSDS_Max-Dir}, is consistent with, e.g., the results of \cite{kristem2018outdoor} for single-antenna systems. 


\begin{table}[h!]
\centering
\caption{Normal fitting for $RMDDS_{Omni}$ with 95$\%$ confidence intervals.}
\label{tab:nf_RMDDS_Omni}
\begin{adjustbox}{width=8.5cm}
\begin{tabular}{|l||c|c|c|c|c|c|}
\hline
\textbf{Frequency} & $\boldsymbol{\mu_{min,95\%}}$ & $\boldsymbol{\mu}$ & $\boldsymbol{\mu_{max,95\%}}$ & $\boldsymbol{\sigma_{min,95\%}}$ & $\boldsymbol{\sigma}$ & $\boldsymbol{\sigma_{max,95\%}}$ \\ \hline \hline
All Bands & -95.99 & -83.53 & -71.06 & 7.41 & 11.88 & 29.13 \\ 
6-7 GHz & -90.96 & -79.87 & -68.78 & 6.60 & 10.57 & 25.92 \\ 
7-8 GHz & -91.02 & -79.92 & -68.82 & 6.60 & 10.58 & 25.94 \\ 
8-9 GHz & -90.85 & -79.81 & -68.76 & 6.57 & 10.52 & 25.81 \\ 
9-10 GHz & -90.59 & -79.75 & -68.90 & 6.45 & 10.33 & 25.34 \\ 
10-11 GHz & -90.26 & -79.45 & -68.64 & 6.43 & 10.30 & 25.27 \\ 
11-12 GHz & -90.49 & -79.49 & -68.50 & 6.54 & 10.48 & 25.70 \\ 
12-13 GHz & -90.31 & -79.29 & -68.26 & 6.56 & 10.50 & 25.76 \\ 
13-14 GHz & -90.65 & -79.81 & -68.98 & 6.45 & 10.33 & 25.32 \\ 
14-15 GHz & -90.27 & -79.57 & -68.87 & 6.37 & 10.20 & 25.01 \\ 
15-16 GHz & -90.74 & -80.03 & -69.31 & 6.37 & 10.21 & 25.05 \\ 
16-17 GHz & -90.70 & -79.97 & -69.24 & 6.38 & 10.23 & 25.08 \\ 
17-18 GHz & -91.17 & -80.38 & -69.60 & 6.41 & 10.28 & 25.20 \\  
\hline
\end{tabular}
\end{adjustbox}
\vspace{-3mm}
\end{table}

\begin{table}[h!]
\centering
\caption{Normal fitting for $RMSDS_{Max-Dir}$ with 95$\%$ confidence intervals.}
\label{tab:nf_RMSDS_Max-Dir}
\begin{adjustbox}{width=8.5cm}
\begin{tabular}{|l||c|c|c|c|c|c|}
\hline
\textbf{Frequency} & $\boldsymbol{\mu_{min,95\%}}$ & $\boldsymbol{\mu}$ & $\boldsymbol{\mu_{max,95\%}}$ & $\boldsymbol{\sigma_{min,95\%}}$ & $\boldsymbol{\sigma}$ & $\boldsymbol{\sigma_{max,95\%}}$ \\ \hline \hline
All Bands & -99.65 & -97.14 & -94.63 & 1.50 & 2.40 & 5.87 \\ 
6-7 GHz & -99.45 & -87.27 & -75.09 & 7.25 & 11.61 & 28.47 \\ 
7-8 GHz & -99.57 & -87.34 & -75.11 & 7.27 & 11.65 & 28.58 \\ 
8-9 GHz & -99.64 & -87.31 & -74.97 & 7.34 & 11.75 & 28.83 \\ 
9-10 GHz & -99.01 & -87.00 & -74.98 & 7.15 & 11.45 & 28.08 \\ 
10-11 GHz & -99.48 & -87.19 & -74.90 & 7.31 & 11.71 & 28.72 \\ 
11-12 GHz & -99.40 & -87.13 & -74.86 & 7.30 & 11.69 & 28.67 \\ 
12-13 GHz & -99.67 & -87.29 & -74.91 & 7.36 & 11.80 & 28.94 \\ 
13-14 GHz & -99.64 & -87.38 & -75.11 & 7.29 & 11.69 & 28.66 \\ 
14-15 GHz & -99.49 & -87.35 & -75.20 & 7.22 & 11.57 & 28.38 \\ 
15-16 GHz & -99.76 & -87.49 & -75.22 & 7.30 & 11.69 & 28.67 \\ 
16-17 GHz & -99.89 & -87.45 & -75.01 & 7.40 & 11.85 & 29.08 \\ 
17-18 GHz & -99.82 & -87.39 & -74.97 & 7.39 & 11.84 & 29.04 \\ 
\hline
\end{tabular}
\end{adjustbox}
\end{table}

\subsection{Angular spread}
Before we analyze the AS, we point out the definition we use, namely Fleury's (see Sec. III), results in AS between 0 and 1, but that for small $\sigma^\circ$ values (up to 0.3), it corresponds to angular spread in radians. We also note the lower bound of the measured AS follows from the beamwidth of the directional antenna. 

It is expected that the Tx angular spread should be small, both because the Tx only scans a $120 ^{\circ}$ sector, and also because the buildings in the scenario form a "street canyon" that concentrates energy in a specific angular range. This is confirmed in Table \ref{tab:nf_AS_TX}. We also note that along the different 1 GHz window bands, AS is not affected by the frequency by much and thus shows similar results for all bands.

\begin{figure}[t!]
	\centering
	\includegraphics[width=0.75\linewidth]{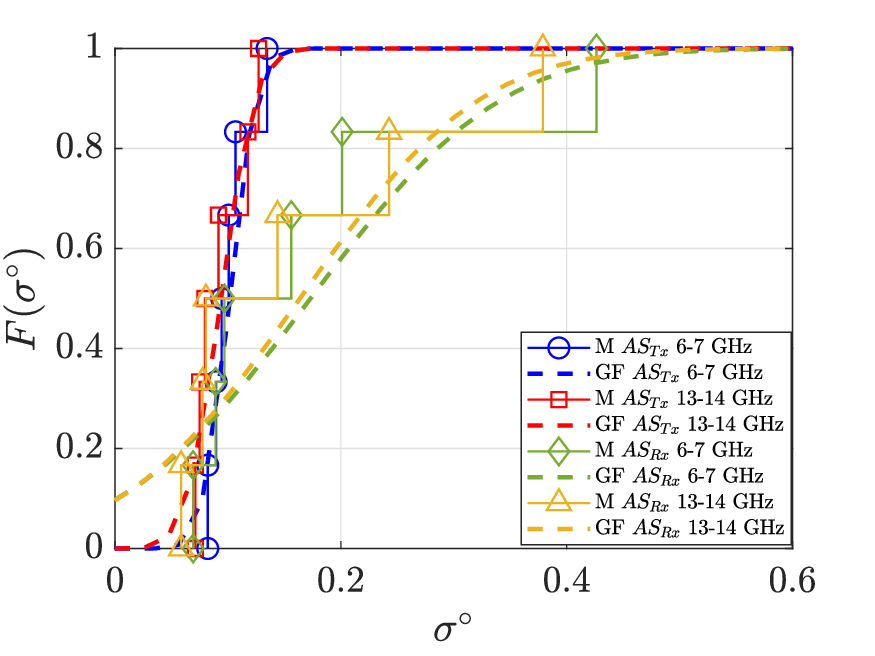}
	\caption{Modeling of AS. Measured (M), and Gaussian Fit (GF) curves.}
	\label{fig:AS}%
 \vspace{-5mm}
\end{figure} 

\begin{table}[h!]
\centering
\caption{Normal fitting for $AS_{Tx}$ with 95$\%$ confidence intervals.}
\label{tab:nf_AS_TX}
\begin{adjustbox}{width=8.5cm}
\begin{tabular}{|l||c|c|c|c|c|c|}
\hline
\textbf{Frequency} & $\boldsymbol{\mu_{min,95\%}}$ & $\boldsymbol{\mu}$ & $\boldsymbol{\mu_{max,95\%}}$ & $\boldsymbol{\sigma_{min,95\%}}$ & $\boldsymbol{\sigma}$ & $\boldsymbol{\sigma_{max,95\%}}$ \\ \hline \hline
All Bands & 0.08 & 0.10 & 0.12 & 0.01 & 0.02 & 0.05 \\ 
6-7 GHz & 0.08 & 0.10 & 0.12 & 0.01 & 0.02 & 0.05 \\ 
7-8 GHz & 0.08 & 0.10 & 0.12 & 0.01 & 0.02 & 0.04 \\ 
8-9 GHz & 0.08 & 0.10 & 0.12 & 0.01 & 0.02 & 0.05 \\ 
9-10 GHz & 0.08 & 0.10 & 0.12 & 0.01 & 0.02 & 0.05 \\ 
10-11 GHz & 0.07 & 0.10 & 0.13 & 0.02 & 0.02 & 0.06 \\ 
11-12 GHz & 0.08 & 0.10 & 0.12 & 0.01 & 0.02 & 0.05 \\ 
12-13 GHz & 0.08 & 0.10 & 0.13 & 0.02 & 0.02 & 0.06 \\ 
13-14 GHz & 0.07 & 0.09 & 0.12 & 0.01 & 0.02 & 0.06 \\ 
14-15 GHz & 0.06 & 0.09 & 0.12 & 0.02 & 0.03 & 0.07 \\ 
15-16 GHz & 0.05 & 0.08 & 0.12 & 0.02 & 0.03 & 0.07 \\ 
16-17 GHz & 0.05 & 0.08 & 0.11 & 0.02 & 0.03 & 0.08 \\ 
17-18 GHz & 0.05 & 0.08 & 0.11 & 0.02 & 0.03 & 0.07 \\
\hline
\end{tabular}
\end{adjustbox}
\vspace{-3mm}
\end{table}

On the other hand, the Rx is expected to have a larger angular spread, given that the scans cover the full azimuthal range and cover all the reflected paths reaching Rx. Moreover, the LoS multipath component has a considerable weight and thus reduces the overall angular spread considerably. In Table \ref{tab:nf_AS_RX-AZ}, we observe statistical fitting for each 1 GHz sub-band and the full UWB case. It is clearly distinguished again that the change in frequency does not have a significant impact on the angular spread, however, the mean value observed in Table \ref{tab:nf_AS_RX-AZ} is approximately double compared to the angular spread of Tx and the variance is at least four times greater than the counterpart of Tx, indicating a more spread power angular spectrum.

Fig. \ref{fig:AS} provides a visual comparison of AS between the 6-7 GHz and 13-14 GHz bands for normal distribution fitting. While the AS values at Tx and Rx are similar for these two frequency bands, both sides have a significant difference between them because of their coverage, in line with previous observations.

\begin{table}[h!]
\centering
\vspace{-2mm}
\caption{Normal fitting for $AS_{Rx}$ with 95$\%$ confidence intervals.}
\label{tab:nf_AS_RX-AZ}
\begin{adjustbox}{width=8.5cm}
\begin{tabular}{|l||c|c|c|c|c|c|}
\hline
\textbf{Frequency} & $\boldsymbol{\mu_{min,95\%}}$ & $\boldsymbol{\mu}$ & $\boldsymbol{\mu_{max,95\%}}$ & $\boldsymbol{\sigma_{min,95\%}}$ & $\boldsymbol{\sigma}$ & $\boldsymbol{\sigma_{max,95\%}}$ \\ \hline \hline
All Bands & 0.02 & 0.18 & 0.34 & 0.10 & 0.15 & 0.38 \\ 
6-7 GHz & 0.03 & 0.17 & 0.31 & 0.08 & 0.13 & 0.33 \\ 
7-8 GHz & 0.02 & 0.18 & 0.34 & 0.09 & 0.15 & 0.37 \\ 
8-9 GHz & 0.02 & 0.18 & 0.35 & 0.10 & 0.16 & 0.38 \\ 
9-10 GHz & 0.03 & 0.17 & 0.31 & 0.08 & 0.13 & 0.32 \\ 
10-11 GHz & 0.04 & 0.17 & 0.31 & 0.08 & 0.13 & 0.32 \\ 
11-12 GHz & 0.02 & 0.18 & 0.33 & 0.09 & 0.15 & 0.36 \\ 
12-13 GHz & 0.02 & 0.18 & 0.34 & 0.09 & 0.15 & 0.37 \\ 
13-14 GHz & 0.03 & 0.16 & 0.30 & 0.08 & 0.13 & 0.31 \\ 
14-15 GHz & -0.00 & 0.18 & 0.37 & 0.11 & 0.17 & 0.43 \\ 
15-16 GHz & -0.00 & 0.18 & 0.36 & 0.11 & 0.17 & 0.43 \\ 
16-17 GHz & -0.01 & 0.17 & 0.35 & 0.11 & 0.17 & 0.42 \\ 
17-18 GHz & 0.02 & 0.15 & 0.28 & 0.08 & 0.12 & 0.30 \\ 
\hline
\end{tabular}
\end{adjustbox}
\vspace{-4mm}
\end{table}

\section{Conclusions}

In this paper, we present an analysis of UWB double-directionally resolved channel measurements in the upper mid-band (FR3). Measurements were conducted at selected points within LoS microcellular scenarios and our analysis reveals that path loss for both Max-Dir and Omnidirectional power delay profiles is lower than the expected free-space values, due to the "street canyon" effect created by surrounding buildings. For RMS delay spread and angular spread, the statistical values remain consistent across different 1 GHz frequency sub-bands, with minimal variation across the entire UWB range, indicating stability in the scenario's multipath propagation characteristics. A similar trend is observed in the angular spread for both Tx and Rx, where the mean value is consistent across all 1 GHz sub-bands. 
These findings provide valuable initial insights into channel characteristics in the FR3 mid-band and contribute to optimizing the design of wireless communication systems.


\end{document}